\documentstyle[12pt,aaspp4]{article}

\newcommand {\hii}{H {\small II}~}
\newcommand {\ha}{H$\alpha$~}
\newcommand {\flux}{ergs^{-1}cm^{-2}~}
\newcommand {\lum}{ergs^{-1}}
\newcommand{\kms}{\ifmmode {\rm \, km \, s^{-1}} \else $\rm \,km, s^{-1}$\fi}

\newcommand{\etal}{{\it et al.}~}

\begin{document}

\title{The Discovery of a Giant \ha  Filament in NGC 7213}
\author{Salman Hameed\altaffilmark{1}}
\affil{Astronomy Department, New Mexico State University, Las Cruces, NM 88003}
\altaffiltext{1}{Visiting Astronomer, Cerro Tololo Inter-American Observatory.
CTIO is operated by the Association of Universities for Research in Astronomy, 
Inc. (AURA) under cooperative agreement with the National Science Foundation.}
\authoremail{shameed@nmsu.edu}
\author{David L. Blank\altaffilmark{2}}
\affil{Department of Astrophysics, School of Physics, University of Sydney, Sydney, NSW 
2006, Australia; Australia Telescope National Facility, CSIRO, P.O. Box 76, Epping, NSW, 2121, 
Australia}
\altaffiltext{2}{The Australia Telescope Compact Array is part of the
of the Australian Telescope which is funded by the Commonwealth
of Australia for operation as a National Facility managed by CSIRO.}
\authoremail{dblank@Physics.usyd.edu.au}

\author{Lisa M. Young}
\affil{Department of Physics, New Mexico Institute of Mining and Technology, 
Socorro, NM 87801}
\authoremail{lyoung@aoc.nrao.edu}
\author{Nick Devereux\altaffilmark{1}}
\affil{Department of Physics, Embry-Riddle Aeronautical University, Prescott, AZ 86301}
\authoremail{devereux@pr.erau.edu}

\begin{abstract}

The nearby Seyfert galaxy NGC 7213 has been imaged in \ha and HI with
the CTIO 1.5 m telescope and with the Australia Telescope Compact Array
(ATCA), respectively. Optically NGC 7213 looks undisturbed and relatively featureless but
the continuum-subtracted \ha image shows a 19 kpc long filament located
approximately 18.6 kpc from the nucleus. The \ha filament could be
neutral gas photo-ionized by the active nucleus, as has been suggested for
the Seyfert galaxy NGC 5252, or shock-ionized by a jet interacting with 
the surrounding HI, as has been suggested for the radio galaxy PKS 2240-41. 
The HI map reveals NGC 7213 to be a highly disturbed system suggesting a past merging event. 

\end{abstract}


\section{Introduction}
NGC 7213 is a face-on Sa (Tully 1988) galaxy located at a distance of 
22.0 Mpc ($H_{0}=75 kms^{-1}Mpc^{-1}$) (\markcite{Tully1988}Tully 1988) and hosts an 
active nucleus.  Its nuclear activity was first discovered in follow-up
optical spectroscopy of X-ray sources observed by the HEAO-A2  satellite
(\markcite{Phillips1979}Phillips 1978) and is in the \markcite{Piccinotti1982}Piccinotti \etal (1982) 
sample of X-ray selected active galaxies. 

The nucleus of NGC 7213 has been classified as a Seyfert 1 (\markcite{Phillips1979}Phillips 1979,
\markcite{FH1984}Filippenko and Halpern 1984)
based on the broad optical emission lines, strong blue continuum and emission
lines from such highly ionized species as Ne$^{+4}$. Its nuclear spectrum 
also exhibits narrow low ionization emission lines that define the LINER class
which may originate from diffuse ionized gas surrounding the nucleus. This 
extended nuclear emission line region (ENER) has been observed in 
several spirals (e.g. M81,M31, N1398 etc.) and its ionization may 
be attributed to shock heating or photo-ionization by the bulge post asymptotic 
giant branch (PAGB) stars (\markcite{Heckman1996}Heckman 1996; 
\markcite{Devereux1997}Devereux, Ford, and Jacoby 1997). 

The optical continuum 
image of the galaxy is dominated by an almost featureless bulge and 
shows no sign of any recent interaction (see Figure 5 of \markcite{HD1999}Hameed
\& Devereux 1999).  An unsharp masking of the 
continuum image shows dust lanes sinuously winding into the 
central region which are also visible in the photograph of
NGC 7213 in the {\it Carnegie Atlas of Galaxies} (\markcite{SB1994}Sandage and Bedke 1994). 
The continuum-subtracted \ha image shows 
a ring of \hii regions surrounding the nucleus that has been studied 
in detail by \markcite{Storchi1996}Storchi-Bergmann \etal (1996). However, the present
paper is motivated by the discovery of a giant \ha filament located approximately 18.6 kpc
(in projection) south of the nucleus. This filament is roughly 
19 kpc long, has no counterpart in the optical continuum image and 
lies well outside the optical diameter of the galaxy (\markcite{HD1999}Hameed
\& Devereux 1999).

An HI map of NGC 7213, obtained as part of an HI survey of nearby 
Seyfert galaxies (\markcite{Blank1999}Blank  1999), reveals NGC 7213 to be a 
highly disturbed system with tidal tails. The \ha filament 
is, in fact, a small part of an HI tail located south of the nucleus. Contrary to the 
optical continuum image, the disturbed HI morphology suggests
that NGC 7213 has gone through a merger or an interaction. In this paper we present 
details of neutral and ionized hydrogen gas morphology of NGC 7213.

\section{Observations}
\subsection{\ha Observations}

An \ha image of NGC 7213 was obtained  on October 25, 1997, using the 
Cassegrain-Focus 
Imager (CFCCD) on the CTIO 1.5m telescope. CFCCD uses a $2048 \times 2048$ 
Tektronics chip and
has a pixel scale of $0.43\arcsec pixel^{-1}$ at f/7.5, yielding a field of view
of $14.7\arcmin \times 14.7\arcmin$. The galaxy was imaged using the 
narrow band \ha + [NII] filter at 6606\AA
($\Delta\lambda$ = 75\AA) and a narrow band line free continuum 
filter at 6477\AA ($\Delta\lambda$=75\AA). Three exposures 
of 900 seconds each were obtained through the line and the continuum filters.
Details of the data reduction process and calibration are described  in 
\markcite{HD2000}Hameed \& Devereux (1999).

\subsection{HI Observations}

NGC 7213 was observed in the 21cm, HI spectral line with the Australia
Telescope Compact Array (ATCA) in its
1.5 D, 750 D, and 750 C configurations on October 9, 1993, and October 
2 and 15, 1995, respectively. The shortest and longest baselines
used were 31 m and 1454 m respectively which resulted in a synthesized beam size
of $42 \arcsec \times 52 \arcsec$. Further details of the HI observations will
be presented in a later paper.

\section{Results}

Figure 1 shows contours of  continuum-subtracted \ha image of NGC 7213 overlayed
on grey-scale HI map of the same region. Despite a 
relatively short integration time (900 sec), the \ha image reveals remarkable 
structure in the ionized gas. The giant filament is located $2.9\arcmin$, or 
18.6 kpc  south of the nucleus (in projection). In comparison, the 
diameter ($D_{25}$) of the galaxy extends out to only $2.1\arcmin$ 
(\markcite{Tully1988}Tully 1988).
The \ha filament 
is roughly 19.0 kpc long ($3.0\arcmin$) and has a total \ha flux of
$F_{H\alpha}=1.1(\pm 0.2) \times 10^{-13}\flux$, which corresponds to 
$L_{H\alpha}=6.4 \times 10^{39}\lum$ at our assumed distance
of 22 Mpc (Tully 1988). The total \ha luminosity of NGC 7213 is 
$L_{H\alpha}=1.7 \times 10^{41}\lum$ (\markcite{HD1999}Hameed \& Devereux 1999).

NGC 7213 has been observed in \ha before. \markcite{Storchi1996}Storchi-Bergmann \etal (1996)
obtained \ha imaging and spectroscopy of the circumnuclear \hii 
regions, but the giant filament was located outside of their field 
of view.  The filament is, however, barely  visible in the \ha image of NGC 7213 obtained
by \markcite{Evans1996}Evans \etal (1996). 

The grey-scale map in Figure 1  shows the neutral gas morphology and 
reveals NGC 7213 to be a highly disturbed system.
The central region has a high concentration of HI and the giant \ha filament 
appears to be a small part of an HI tail located southwest of the nucleus. 
This HI tail has a peak surface density of $5.1 M_{\odot}pc^{-2}$.
There appears to be a lack of HI emission in a region 
that is $40\arcsec \times 60\arcsec$ in size lying about $2\arcmin$
southwest of the nucleus and is not due to HI absorption (For details, 
see \markcite{Blank1999}Blank 1999).

Figure 2 shows a close-up of the \ha filament and the region surrounding 
it. Overall the filament has a bow-like structure with a discontinuity 
towards the eastern end. The filament does not have any counterpart in 
the optical continuum image (\markcite{HD1999}Hameed \& Devereux 1999). There is some low level 
diffuse \ha emission that connects the filament with a ring of \hii  regions surrounding the 
nucleus.

Figure 3 reveals the total extent of neutral gas which extends far beyond
the optical radius. The azimuthally averaged column density falls to 
$1 \times 10^{19}cm^{-2}$ at a radius of about 9 arcminutes (58 kpc),
and the azimuthally averaged surface density of HI is $1 M_{\odot}pc^{-2}$ 
at a radius of 3.5 arcminutes. The overall HI distribution is clumpy and roughly 
oval in shape with a position angle of about $110^{\circ}$. There is a 
prominent HI tidal tail in the north-western region of
the galaxy. The total HI mass of the NGC 7213 system is 
$4.6 \times 10^{9}$ $M_{\odot}$.  

The velocity field of NGC 7213 is shown in Figure 4. The velocity 
field becomes increasingly disordered further away from the center and 
is likely to be the result of a merger. 

\section{Discussion}

{\it What is the source of ionization for the giant filament?}
It is unlikely that the filament is foreground galactic emission since 
our \ha image was obtained with a redshifted \ha filter and any foreground emission
would have been subtracted out. In addition,  the \ha morphology and its 
correlation with HI strongly suggests a connection with NGC 7213.
The lack of clumpy HII regions in the filament also rule out local star formation as the 
ionization source. Similarly, there is no continuum emission from the 
filament, ruling out field OB and post-asymptotic Giant Branch stars as a 
source of ionization. 

There are at least two possible mechanisms which might 
explain the large \ha filament. One possibility is a starburst-related
superwind. \markcite{DB1999}Devine \& Bally (1999) found
a 3 kpc long ionized ``cap'' located 11 kpc above the plane of M82. 
Fainter \ha was also detected between the \ha cap and M82. Devine and
Bally have suggested that the \ha in the cap traces ambient material that 
is being shocked by a superwind and/or photoionized by radiation
from nuclear starburst. Massive star formation in NGC 7213 is confined to a ring 
of HII regions surrounding the nucleus and is much weaker than the starburst in M82. 
Our calculations suggest that it is very unlikely that the filament in 
NGC 7213, which is much larger than the ``cap'' in M82(Table 1), is being 
ionized by the leakage of UV photons from the star forming ring.

A more likely explanation is that the filament is either being photo-ionized 
by UV photons escaping from the central AGN or shock-ionized by a jet 
interacting with the surrounding medium. Observations of Seyfert galaxies 
have identified galactic scale minor outflows in a number of 
galaxies (e.g. \markcite{WT1994}Wilson \& Tsvetanov 1994; 
\markcite{Colbert1996}Colbert \etal. 1996).
The presence of broad emission lines (\markcite{FH1984}Filippenko \& Halpern 1984), along with the 
detection of X-rays(\markcite{Piccinotti1982}Piccinotti et al. 1982; 
\markcite{Fabbiano1992}Fabbiano, Kim, and Trinchieri 1992) and an 
unresolved UV point source (\markcite{RB1999}Rokaki \& Boisson 1999) in the nucleus of NGC 7213,  
strongly suggest  the presence of an AGN. 
Lyman continuum photons from the central engine may ionize part of the 
extensive HI which surrounds NGC 7213. Such a situation exists within the type 1 
Seyfert galaxy NGC 5252, which hosts a biconical structure of ionized gas that 
extends up to a distance of 18 kpc from the nucleus and consists of 
a complex network of filaments(\markcite{PF1996}Prieto \& Freudling 1996; 
\markcite{Tsvetanov1996}Tsvetanov \etal 1996; 
\markcite{Morse1998}Morse \etal 1998). The largest of these filaments 
is located about 7 kpc from the nucleus and is approximately 3.6 kpc in 
length (Table 1). 

In the case of NGC 7213, the \ha luminosity of the nucleus 
is $4.7 \times 10^{40} \lum$. If the filament in NGC 7213 is photoionized 
by the nucleus, then, at least, 14$\%$ of ionizing photons have to escape
from the nucleus to provide the observed \ha luminosity of the filament:
$6.4 \times 10^{39} \lum$. We can obtain a rough estimate as to whether or not 
the source responsible for \ha emission in the nucleus has enough 
energy to ionize the filament. Lets consider a simple case where ionizing 
photons are leaking isotropically from the nucleus. The flux at the distance of
of the filament (18.6 kpc), $F_{nf}$, can be calculated by $F_{nf}=L(H\alpha)_{nuc}/4\pi D^{2}$,
where $L(H\alpha)_{nuc}$ is \ha luminosity of the nucleus  and D is the distance of the
filament from the nucleus. Using the above values, we get $F_{nf}=1.1 \times 10^{-6} \flux$. 
Assuming the filament to be a 2-dimensional sheet, we can get the luminosity of the 
filament, $L_{nf}$, by multiplying $F_{nf}$ by the surface area of the filament, 
$165 \arcsec \times 55 \arcsec$ or $5.9 \times 10^{22}cm \times 1.8 \times 10^{22}cm$.
This gives $L_{nf} = 1.2 \times 10^{39} \lum$ which is 
about 6 times smaller than the observed luminosity of the filament (Table 1). It should
be noted, however, that the \ha luminosity of the nucleus is not an accurate measure 
of the ionizing flux, and represents only those ionizing photons that have been 
absorbed by hydrogen near the nucleus. The covering factor for the broad line 
region in AGNs is quite uncertain, but some studies (e.g. \markcite{Peterson1997}Peterson 1997)
estimate the covering factor to be around 0.1, suggesting that only 10$\%$ of the
ionizing photons are being absorbed. If this is the case in NGC 7213, then 
the nucleus can easily provide the photons required to ionize the filament.
In addition, our calculations above used the simplest case of
isotropic emission. Even a modest beaming of the ionizing 
photons towards the filament can potentially provide enough energy to ionize
the filament.

On the other hand, the radio galaxy PKS 2250-41 provides an example where a giant 
emission line arc is shock ionized by a radio jet(\markcite{Tadhunter1994}Tadhunter \etal 1994). 
It should be noted, however, that the emission line arc in PKS 2250-41 has been observed
only in [OIII] and no \ha observations for the galaxy exist in the literature. The arc in 
PKS 2250-41 is located 37 kpc from the nucleus and has a total length 
of about 50 kpc (\markcite{Clark1997}Clark \etal 1997). 
Such jet interaction induced ionization could be the 
case in NGC 7213, albeit on a smaller scale. In the 
study of star forming regions of NGC 7213, 
\markcite{Storchi1996}Storchi-Bergmann \etal (1996) found 
velocity dispersions near the nucleus suggesting a collimated outflow or 
other non-circular motions. Furthermore, 
\markcite{Harnett1987}Harnett (1987) and \markcite{Blank1999}Blank (1999), found extended 
radio emission in low resolution images that might also indicate the presence of a radio jet.

Understanding the kinematics of the ionized gas could help distinguish
between these two possibilities. If ionized gas is produced by
UV photons escaping from the nucleus, then we might expect the ionized gas
to have the same radial velocity as the neutral gas at the same location. On the 
other hand, if the ionized gas is produced by shocks, its velocity should be 
significantly different from the neutral gas. In addition, optical spectroscopy can also 
provide diagnostic line ratios that may determine whether the filament 
is shock-ionized or photo-ionized. Shock-ionization is expected to produce high ratios 
of [NII](6548+6584$\AA$)/\ha,[SII](6716+6731$\AA$)/\ha, and [OIII](5007$\AA$)/\ha
relative to photo-ionization (\markcite{SM1979}Shull \& McKee 1979;
\markcite{DS1995}Dopita \& Sutherland 1995).

The identity of the progenitor galaxy, responsible for the disturbed HI morphology
of NGC 7213, is also unclear.  There is a nearby elliptical galaxy of unknown 
redshift located at $5\arcmin$ north-east of NGC 7213. If that system 
is a close companion, it is unlikely to be responsible for the disturbed 
HI morphology. While the anomalous HI arms  could be produced by a small galaxy 
tidally removing HI gas from its larger companion (e.g. the M 51 system), the 
elliptical companion of NGC 7213 is located in the middle of the northern arm, not at 
the end of the arm as one would expect, if it was drawing out the HI. It is 
more likely that 
NGC 7213 is a merger remnant, as \markcite{Blank1999}Blank (1999) has identified 
an inner HI disk counter-rotating with respect to the larger HI disk in 
NGC 7213. This notion is further supported by the presence of 
``shells'' in an unsharp mask of the Digital Sky Survey image of NGC 7213
(Hibbard, private communication). However, we cannot rule out the presence of another 
perturber, as there is a bright foreground star to the north-west of NGC 7213 that may 
be masking any nearby galaxy.

The giant filament in NGC 7213, one of the largest 
among nearby active galaxies, joins an increasing number of galaxies
that have galactic scale ionization features. It is difficult to 
estimate how common these objects are, as very few studies of 
Seyfert galaxies exist that have focused on large scale features. 
In a study of 22 edge-on Seyfert galaxies, \markcite{Colbert1996}Colbert et al. (1996) find 
$\sim 1/4$ of galaxies with good evidence for minor axis outflows, but none
at scales comparable to NGC 7213 and NGC 5252. Similarly, \markcite{Baum1993}Baum et al. (1993)
found kpc-scale extended features in seven out of ten Seyfert galaxies they studied.
A large, wide field, systematic study of nearby Seyfert galaxies will provide 
information about the importance of these filaments and their relationship 
to AGNs and their host galaxies.

\acknowledgments
S.H. would like to thank NOAO for providing travel support to 
CTIO and Kurt Anderson, John Hibbard, and Charles Hoopes for useful comments on 
the paper. DLB would like to thank the University of Sydney and the Australian
Department of Employment, Education and Training for respectively an
overseas postgraduate research scholarship and for an overseas postgraduate
research award. DLB also thanks the ATNF for additional funding and for the
generous use of its facilities. The authors would also like to thank the
referee for useful comments that improved the presentation of the paper.

\clearpage

\begin{deluxetable}{lcccccc}
\scriptsize
\tablenum{1}
\tablewidth{0pt}
\tablecaption{Comparison of Filaments in 3 galaxies}
\tablehead{
\colhead{Galaxy} & \colhead{Filament size} & \colhead{Distance from 
the nucleus} & 
\colhead{H$\alpha$ luminosity} & \colhead{Possible ionization source} 
& \colhead{Reference}\\
& \colhead{(kpc)} & \colhead{(kpc)} & \colhead{($\lum$)}& &
}
\startdata
M 82  & 3    & 11 & $2 \times 10^{38}$ & Starburst wind &  
Devine \& Bally (1999)  \\
NGC 5252  &  3.6   & 7 & $5.3 \times 10^{39}$ & Active nucleus & Morse et al. (1998) \\
NGC 7213  & 19.0    & 18.6 & $6.4 \times 10^{39}$ & Active nucleus & This paper \\


\enddata
\end{deluxetable}

\clearpage

\figcaption{Contours of Continuum-subtracted \ha image of NGC 7213 overlayed on  
HI map of the same region. North is at the top and east is to the left 
of the image. The giant \ha filament is located on 
the southern HI tail located $3\arcmin$ south of the nucleus. The contours are
13, 26, 40, 66, 198, 594, and 1782 Emission Measure (pc cm$^{-6}$). Assuming a distance of 
22.0 Mpc, $1 kpc = 9.4\arcsec$. Coordinates in the image are accurate to $\sim$2 arcsec. 
The ellipse in the lower-left corner of the HI image shows the resolution(FWHM) of that image.}

\figcaption{Image showing the complicated structure of the filament and other ionized 
features near the nuclear region. The black bar in the bottom left corner of the image
represents 1 kpc in length.}

\figcaption{HI contours overlaid on grey-scale of integrated HI distribution. The 
contours are 10,20,30,40,50,60,70,80,90 percent of the peak, which is
$6.6 \times 10^{20}$ H atoms cm$^{-2}$. Note that the optical image of NGC 7213
would cover just the central $2.1\arcmin$ of the HI map.}

\figcaption{The isovelocity curves. The contours are from 1508 to 
2028 $km s^{-1}$ in intervals of 40 $km s^{-1}$.}

\end{document}